physica status solidi

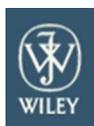
physica status solidi

# Photoinduced current in molecular conduction junctions with semiconductor contacts

# Photoinduced current in molecular conduction junctions with semiconductor contacts

**Boris Fainberg** [*,1,2] **and Tamar Seideman** [3]

[1] Faculty of Science, Holon Institute of Technology, Golomb street 52, 58102 Holon, Israel
[2] School of Chemistry, Tel-Aviv University, 69978 Tel-Aviv, Israel
[3] Department of Chemistry, Northwestern University, 60208 Evanston, IL, USA

We propose a new approach to coherent control of transport via molecular junctions, which bypasses several of the hurdles to experimental realization of optically manipulated nanoelectronics noted in the previous literature. The method is based on the application of intrinsic semiconductor contacts and optical frequencies below the semiconductor bandgap. It relies on a simple and general concept, namely the controllable photonic replication of molecular levels through the dipole driving the molecular bridge by an electromagnetic field. We predict the effect of coherent destruction of induced tunneling that extends the certain effect of coherent destruction of tunneling. Our results illustrate the potential of semiconductor contacts in coherent control of photocurrent.

**1 Introduction** The optical response of nanoscale molecular junctions has been the topic of growing experimental and theoretical [1-5] interest in recent years, fueled by both the rapid advance of the experimental technology and the premise for long range applications. The ultimate goal of controlling electric transport with coherent light, however, has proven challenging to realize in the laboratory. One difficulty that has been noted in the previous literature is substrate mediated processes. Light shine on a molecular system in contact with a metal substrate is absorbed by the substrate, rather than by a molecular bond or the molecule-surface bond in the vast majority of cases, leading to the excitation of hot carriers. The latter may interact with the molecule resulting in the loss of coherence. Other competing processes include heating of the electrodes and undesired energy transfer events. Reference [1] proposes the use of semiconducting (SC) electrodes and sub-bandgap frequencies to circumvent undesired substrate-mediated and heating processes. Here, an ultrafast, nano-scale molecular switch is introduced, consisting of a conjugated organic molecule adsorbed onto a SC surface and placed near a scanning tunneling microscope tip. A low-frequency, polarized laser field is used to switch the system by orienting the molecule with the field polarization axis, enabling conductance. SC electrodes have been used in the experimental literature in the context of a single-quantum-dot photodiode that may be considered as a quantum dot-based junction [6,7]. In addition to introducing a new opportunity for coherent control of transport via junctions, SC-based molecular electronics offer potentially several other attractive properties. From a chemical perspective, organic molecules form much stronger bonds with SC surfaces, such as doped silicon, than with metals. From a technological perspective, the addition of molecular function to the already established silicon-based technology is vastly more viable than replacing silicon by metal-based electronics. Here we propose a new approach to coherent control of electric transport via SC junctions, which is similar to the concept introduced in Ref. [1] in capitalizing on the use of sub-bandgap frequencies, but is complementary in application. Our approach is based on the excitation of dressed states of the junction Hamiltonian that can be frequency-tuned to tunnel selectively into either the left or the right contacts, generating unidirectional current whose temporal characteristics are controlled by the light pulse.

**2 Control concept**
We consider a molecular junction consisting of a molecular moiety that is in contact with two intrinsically semiconducting electrodes. The use of SC contacts circumvents energy transfer from the bridge to the contacts [7], a complicating feature in junctions with metallic contacts, since, as noted above, sub-bandgap light cannot excite electron-hole pairs in a semiconductor substrate [1]. Hence, the main source of relaxation in SC-molecule-SC junctions under the conditions considered is the charge transfer between the bridge and the contacts. If the molecular moiety possesses a permanent dipole moment $\mathbf{D}_{ii}$, the interaction of a nonresonant electromagnetic (EM) field with such systems leads to modulation of their energetic spectrum by the field frequency $\omega$. This modulation alters the arrangement of molecular electronic states and may substantially change the electron transfer rates. The efficiency of the energy spectrum modulation depends on the interaction parameter $z_i = \mathbf{D}_{ii} \cdot \mathbf{E}_0 / (\hbar\omega)$, where $\mathbf{E}_0$ is the amplitude of the EM field $\mathbf{E}(t)$. The permanent dipole moment of relevant molecules can reach 10 $D$ and more. It seems likely that this idea was first applied to the activation of radiationless transitions in large molecules [8,9]. The modulation under discussion may also substantially change the electron and hole transfer rates between the molecular bridge and the SC contact, due to the strong dependence of these rates on the position of a molecular level relative to the conduction band (CB) and valence band (VB). Suppose that a single molecular level of energy $E_i$ is positioned between the conduction and valence bands of the SC contacts shown in Fig.1. No current in such SC-molecule-SC junction is possible even in the presence of the voltage bias. If, however, an EM pulse of appropriate frequency $\omega$ excites the molecular bridge, photonic replication of state $i$ with energy $E_i + \hbar\omega$ can be tuned to be close to the CB, while the photonic replication with energy $E_i - \hbar\omega$ are close to the VB coupling state $i$. In that situation a current flows through the junction and its temporal duration is controlled by the EM pulse characteristics. The transport rate is largely controlled by the applied voltage bias, which determines the barrier width. This control enables us to realize coherent excitation of a molecular bridge while circumventing competing processes. The control concept under discussion is referred to the photon assisted transfer (PAT), a theory of which in nanojunctions with metallic contacts was developed by Haenggi et al. [2,3]. In particular, they considered PAT across a bridged molecular wire. In these works the time dependent level shift in the bridge arises from the coupling to an oscillating dipole field (see below). For a single site bridge this coupling may be described by a permanent dipole moment [5].

### 3 Theory
**3.1 Model** The complete Hamiltonian describing a molecular bridge interacting with two SC electrodes and subject to a low frequency optical pulse is written as,

$$\hat{H} = \hat{H}_{SC} + \hat{H}_{wire} + \hat{W} + \hat{V} \qquad (1)$$

where $\hat{H}_{SC} = \sum_{k \in \{L,R\}} (\varepsilon_{ck} \hat{c}_{ck}^\dagger \hat{c}_{ck} + \varepsilon_{vk} \hat{c}_{vk}^\dagger \hat{c}_{vk})$ is the Hamiltonian of intrinsic semiconductor leads, $\hat{c}_k^+$ ($\hat{c}_k$) ($k \in L,R$) are creation (annihilation) operators for electrons (energies $\varepsilon_{c,vk}$) in the leads, $c$ and $v$ denote the conduction and valence bands, respectively, and $L(R)$ stands for the right (left) lead. In what follows we will omit the band indices $c$ and $v$ when not essential, so as to simplify the notation. The wire Hamiltonian,

$$\hat{H}_{wire} = \sum_{m=1}^{N} E_m \hat{c}_m^\dagger \hat{c}_m - \Delta \sum_{m=1}^{N-1} (\hat{c}_{m+1}^\dagger \hat{c}_m + \hat{c}_m^\dagger \hat{c}_{m+1}) \qquad (2)$$

is described as a tight-binding model composed of $N$ sites, where each site represents available orbitals, $\hat{c}_m^+$ ($\hat{c}_m$) ($m=1,...N$) are creation (annihilation) operators for electrons in the different molecular states of energies $E_m$. The $\Delta$ term in Eq. (2) accounts for electron transfer interactions between nearest sites within the Huckel model, and $\hat{W} = -\mathbf{D} \cdot \mathbf{E}$ describes the interaction of the bridge sites with an external EM field $\mathbf{E}$ where the dipole operator has only diagonal elements $D_{mm} = D(N+1-2m)/2$ with $D$ equal to the electron charge multiplied by the distance between the neighboring sites [2,3]. Finally, $\hat{V} = \sum_{n=1,N; k \in K} (V_{nck} \hat{c}_{ck}^\dagger \hat{c}_n + V_{nvk} \hat{c}_{vk}^\dagger \hat{c}_n) + h.c.$

where $h.c.$ denotes Hermitian conjugate, describes electron transfer between the molecular bridge and the leads, thus giving rise to net current in the biased junction.

**3.2 Equations of motion** Our analysis is based on the generalized master equation for the reduced density matrix of the molecular system taking $\hat{V}$ as a perturbation [2,3, 5,8]. Briefly, one starts with the equation for the total density operator and uses projectors of the type $P_K \rho(t) = \rho_K Tr_K \rho(t)$ in order to derive an equation for the time evolution of the reduced density matrix $\sigma(t) = Tr_R Tr_L \rho(t)$. This leads to

$$\frac{d\sigma}{dt} + \frac{i}{\hbar}[\hat{H}_0, \sigma] = \sum_{nrq} \gamma_n^{rq} \{-\{([\hat{C}_n^q, \hat{C}_n^{\dagger r}] f_K(\hbar\omega_r) + \hat{C}_n^{\dagger r} \hat{C}_n^q), \sigma\}$$
$$+ 2\hat{C}_n^q \sigma \hat{C}_n^{\dagger r} [1 - f_K(\hbar\omega_r)] + 2\hat{C}_n^{\dagger r} \sigma \hat{C}_n^q f_K(\hbar\omega_r)\} \qquad (3)$$

where $\{\hat{A}, \hat{B}\} = \hat{A}\hat{B} + \hat{B}\hat{A}$, $f_K(\hbar\omega_r)$ is the Fermi function,

$$f_K(\hbar\omega_r) = \frac{1}{\exp[(\hbar\omega_r - \mu_{c,v})/k_B T] + 1} \qquad (4)$$

of the CB ($c$) or VB ($v$), $k_B$ is Boltzmann's constant, $T$ is the absolute temperature, and $\mu_{c(v)}$ is the quasichemical potential in the CB (VB).

$$\hat{C}_n^{q,r} = \exp(-i\hat{H}_0 t/\hbar) \hat{c}_n^{q,r} \exp(i\hat{H}_0 t/\hbar), \qquad (5)$$

where $\hat{c}_n^{q,r}$ are the Fourier amplitudes of the interaction picture operators $\hat{c}_m^{+\text{int}}(t)$ and $\hat{c}_m^{\text{int}}(t)$

$$\hat{c}_n^{+\text{int}}(t) = \sum_r \hat{c}_n^{+r}\exp(i\omega_r t), \quad \hat{c}_n^{\text{int}}(t) = \sum_q \hat{c}_n^q \exp(-i\omega_q t) \quad (6)$$

contained in $\hat{V}^{\text{int}}(t) = \exp(i\hat{H}_0 t/\hbar)\hat{V}\exp(-i\hat{H}_0 t/\hbar)$, the interaction representation of $\hat{V}$, $\hat{H}_0 = \hat{H}_{SC} + \hat{H}_{wire} + \hat{W}$

$$\gamma_n^{rq} = \frac{\pi}{\hbar^2}\sum_k |V_{nk}|^2 \exp[i(\omega_r - \omega_q)t]\xi(\omega_r - \omega_q)\delta(\omega_k - \omega_r) \quad (7)$$

is the spectral function, $\omega_k = \varepsilon_k/\hbar$. The frequency dependence of $\gamma_{mn}^{rq}$ may be neglected provided that that $\gamma_{mn}^{rq}$ is small relative to the bath correlation frequency $\omega_c$ - the range over which its spectral density essentially changes. In Eq. (7) we retained only the terms giving the dominant contributions, for which $|\omega_r - \omega_q| \ll \omega_c$. This was done by introducing a switching function $\zeta(\omega_r - \omega_q)$ defined as $\zeta(\omega_r - \omega_q) = 1$ for $|\omega_r - \omega_q| \ll \omega_c$, and $\zeta(\omega_r - \omega_q) = 0$ for $|\omega_r - \omega_q| \geq \omega_c$.

**3.3 An analytically soluble model** We conclude this section by considering briefly the simplest case scenario of a single site bridge. Specifically, we envision a single molecular level of energy $E_i$ (Fig.1) that is excited by an EM field $\mathbf{E}(t) = \mathbf{E}_0 \cos(\omega t)$ tuned to a sub-bandgap frequency $\omega$, such that the dressed energies $E_i + \hbar\omega$ and $E_i - \hbar\omega$ are close to the conduction and valence bands, respectively. One finds for the case under consideration

$$\hat{c}_i^{\text{int}}(t) = \hat{c}_i \exp[-i\omega_i t + iz_i \sin(\omega t)] = \sum_{r=-\infty}^{\infty} \hat{c}_i^r \exp(-i\omega_r t) \quad (8)$$

where $\omega_i = E_i/\hbar$, $\omega_r = \omega_i - r\omega$, $\hat{c}_i^r = \hat{c}_i J_r(z_i)$, $J_r(z_i)$ is the $r$th-order Bessel function. This expansion can be extended to exciting a molecular bridge by pulsed, rather than continuous wave (CW) light, $E(t) = E_0(t)\cos\omega t$ (whose pulse duration is long with respect to the optical cycle [10]). In that case the interaction parameter, defined above, $z_i(t) = \mathbf{D}_{ii} \cdot \mathbf{E}_0(t)/(\hbar\omega)$, acquires a time dependence. The time evolution of the molecular bridge population $n_i = \langle \hat{c}_i^\dagger \hat{c}_i \rangle = Tr(\hat{c}_i^\dagger \hat{c}_i \sigma)$ obtained with the density matrix, Eq. (5), reduces in the limit of a single site bridge, Fig.1, to

$$\frac{dn_i}{dt} = \sum_K [2(1-n_i)\Gamma_{vK,i} - 2n_i\Gamma_{cK,i}] \quad (9)$$

where $\Gamma_{vK,i} = \sum_{r=1}^{\infty} J_r^2(z_i)\gamma_{vK,i}^{rr}$, $\Gamma_{cK,i} = \sum_{r=-\infty}^{-1} J_r^2(z_i)\gamma_{cK,i}^{rr}$, $\gamma_{cK,i}^{rr}$ and $\gamma_{vK,i}^{rr}$ are $\gamma_i^{rr}$ for $k \in cK$ (CB of lead $K$) and for

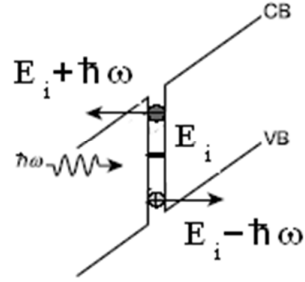

**Figure 1** Biased SC-molecule-SC junction. Electromagnetic excitation applied to a molecular bridge in state $E_i$ leads to generation of its photonic replica at energies $E_i \pm \hbar\omega$. Tunnelling from the photonic replica results in a unidirectional photocurrent.

$k \in vK$ (VB of lead $K$), respectively. For our model, $\gamma_{vL,i}^{11} = \gamma_{cR,i}^{-1,-1} = 0$. The right-hand side of Eq. (11) is proportional to the sum of $I_R(t)$ and $I_L(t)$, where

$$I_K = -2e[(1-n_i)\Gamma_{vK,i} - n_i\Gamma_{cK,i}] \quad (10)$$

- the current due the coupling of state $i$ with electrode $K$. The first order differential equation (9) can be readily integrated, giving for $n_i(0) = 0$, and excitation by a rectangular pulse of duration $t_p$,

$$I_K = \frac{2e}{\Gamma_{vi} + \Gamma_{ci}}\{(\Gamma_{cK,i}\Gamma_{vK',i} - \Gamma_{vK,i}\Gamma_{cK',i}) - (\Gamma_{cK,i} + \Gamma_{vK,i})\Gamma_{vi}\exp[-2(\Gamma_{vi} + \Gamma_{ci})t]\} \quad (11)$$

where $\Gamma_{vi} = \sum_K \Gamma_{vK,i}$ and $\Gamma_{ci} = \sum_K \Gamma_{cK,i}$ are the rates of electron transfer between molecular state $i$ and VB and CB, respectively, of both leads; $K'$ denotes a lead opposite to lead $K$, and we used Eq.(10). Eq.(11) gives for $|z_i| \ll 1$

$$I_R(t) = \frac{-2e\Gamma_R}{\Gamma_L + \Gamma_R}\{\Gamma_L + \Gamma_R \exp[-2(\Gamma_L + \Gamma_R)t]\}, \quad (12)$$

$$I_L(t) = 2e\frac{\Gamma_R\Gamma_L}{\Gamma_L + \Gamma_R}\{1 - \exp[-2(\Gamma_L + \Gamma_R)t]\}, \quad (13)$$

where the dependence on the bias voltage and the laser parameters is implicit in the $\Gamma_R = J_1^2(z_i)\gamma_{vR,i}^{11}$ and $\Gamma_L = J_1^2(z_i)\gamma_{cL,i}^{-1,-1}$. A plot of $-I_R(t)$ and $I_L(t)$ versus time is shown in Fig.2 together with $e(dn_i/dt) = -I_R - I_L$ (see below). The hole current, $-I_R$, starts out at a finite value, $2e\Gamma_R$, and decays exponentially according to the tunneling lifetime $[2(\Gamma_L + \Gamma_R)]^{-1}$, building up negative charge on the bridge. Correspondingly, the electron current, $I_L$, starting at zero, grows at the same rate, the tunnelling rate. In the steady-state regime the values of both currents coincide, $-I_R = I_L = 2e\Gamma_L\Gamma_R/(\Gamma_L + \Gamma_R)$ - the rate of building up negative charge on the bridge is equal to that of its leaving through CB. An observable quantity in pulsed experiments is the charge transferred during an electromagnetic pulse $Q_{L,R} = \int_0^\infty I_{L,R}(t)dt$ [6]. We evaluate a necessary field strength $\sim 10^6$ V/cm, as applied to experimental arrangement [6]. Using Eq.(10), Eq.(9) can be written as $dn_i/dt = -I_R/e - I_L/e$. At short time $dn_i/dt$ is positive (see Fig.2) - the charge density on the bride grows at a constant rate determined by $\Gamma_R$. As electron current develops, the rate of change of $n_i$ drops, until it stabilizes on zero at steady state, where the charge density on the bridge remains constant. For pulses short with respect to the tunneling time, however, steady state is never established and $n_i$ maintains its time dependence throughout. Integrating Eq.(9) from $t=0$ to $t=\infty$, we get, $e[n_i(\infty)-n_i(0)] = -Q_R - Q_L$. The last equation expresses the charge conservation - the transferred charges value $-Q_R$ and $Q_L$ coincide only for $n_i(\infty) = n_i(0)$. This is achieved when the pulse is long with respect to the tunneling lifetime, $t_p \gg [2(\Gamma_L + \Gamma_R)]^{-1}$ for $|z_i| \ll 1$. It is readily seen that current in the SC-molecule-SC junction exists only in the presence of external EM field ($|z_i| \neq 0$) - in the absence of the laser pulse both $I_R$ and $I_L$ vanish. Eqs. (12-13) illustrate also the possibility of generating unidirectional current at times short with respect to $[2(\Gamma_L + \Gamma_R)]^{-1}$.

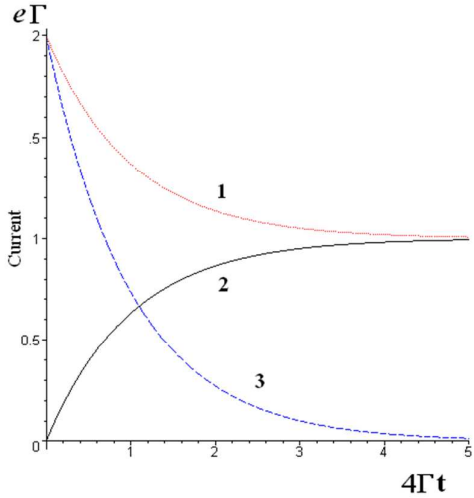

**Figure 2** Currents $-I_R(t)$ (1), $I_L(t)$ (2) and $-I_R(t)-I_L(t)$ (3) versus time for $\Gamma_R = \Gamma_L \equiv \Gamma$ and $|z_i| \ll 1$.

Consider steady-state current, the first term on the right-hand side of Eq.(11), when parameter $|z_i|$ is not small. It can be expected for our model with broad CB and VB that due to symmetry, the corresponding spectral functions of the $r$-th order for the same band are equal for the left and right contacts, i.e. $\gamma_{cL,i}^{rr} = \gamma_{cR,i}^{rr}$ and $\gamma_{vR,i}^{rr} = \gamma_{vL,i}^{rr}$ when $|r|>1$. For this case the steady-state current acquires the form

$$I_L = -I_R = 2eJ_1^2(z_i)\frac{\gamma_{cL,i}^{-1,-1}(\Gamma_{vR,i}-\gamma_{vR,i}^{11})+\gamma_{vR,i}^{11}\Gamma_{cL,i}}{\Gamma_{vi}+\Gamma_{ci}} \quad (14)$$

Eq.(14) shows that the current is zero at zeros of the first-order Bessel function $J_1(z_i)$. As the matter of fact, we obtain the effect of coherent destruction of induced tunneling (CDIT) that extends the certain effect of CDT related to zeros of $J_0(z_i)$ [2,5]. The point is that the coupling of the molecular bridge with semiconductor leads $V_{ic,vk}$ should be substituted by the effective coupling [2] that in our case is equal to $(V_{ic,vk})_{eff} = J_{\mp 1}(z_i)V_{ic,vk}$ for the electron transfer rates $\gamma_{cL,i}^{-1,-1}$ and $\gamma_{vR,i}^{11}$, respectively (see Eq.(14)). The effective tunneling matrix element is suppressed when $J_{\mp 1}(z_i)$ is equal to zero. To build a plot of the steady-state current versus $z_i$, one needs to know $z_i$-dependence of parameters $\Gamma$ in Eq.(14) that implies knowing the $r$-dependence of $\gamma_{c,vKj}^{rr}$, Eq.(7). Using the density of states of a 3D semiconductor, one gets for $|r|=2,3,4...$
$\gamma_{c,vK,j}^{rr} = \sqrt{|r|-1}\gamma_{c,vK,j}^{\mp 2,\mp 2}$. The value of $\gamma_{c,vK,j}^{\mp 1,\mp 1}$ strongly depends on the applied voltage bias (tunnel effect) and may vary between $10^{-2}-10^{-4}eV$.

**4 Conclusion** We proposed a viable approach to coherent control of electric transport via molecular junctions and developed a theoretical framework to explore the method. Our approach makes use of SC electrodes and sub-bandgap frequencies to circumvent substrate-mediated processes and competing energy transfer events. It relies on a simple and general concept, namely the controllable photonic replication of molecular levels through the dipole driving the molecular bridge by an EM field. Importantly, our approach circumvents undesired substrate-mediated effects and energy transfer events.